\documentclass[a4paper,fleqn,usenatbib,letter]{mnras}

\usepackage{mathptmx}

\usepackage[T1]{fontenc}
\usepackage{ae,aecompl}

\usepackage{graphicx}	
\usepackage{amsmath}	
\usepackage{amssymb}	

\title[Magnetic Fields and the Evaporation Valley]{Effects of Magnetic Fields on the Location of the Evaporation Valley for Low-Mass Exoplanets}

\author[Owen, J. E. \& Adams F. C.]{
James E. Owen$^{1}$\thanks{E-mail: james.owen@imperial.ac.uk} and
Fred. C. Adams$^{2,3}$
\\
$^{1}$Astrophysics Group, Blackett Laboratory, Imperial College London, Prince Consort Road, London, SW7 2AZ, UK\\
$^{2}$Physics Department, University of Michigan, Ann Arbor, MI 48109, USA \\
$^{3}$Astronomy Department, University of Michigan, Ann Arbor, MI 48109, USA
}

\pubyear{2015}

\begin{document}
\label{firstpage}
\pagerange{\pageref{firstpage}--\pageref{lastpage}}
\maketitle

\begin{abstract}
The observed distribution of radii for exoplanets shows a bimodal form that can be explained by mass-loss from planetary atmospheres due to high energy radiation emitted by their host stars. The location of the minimum of this radius distribution depends on the mass-radius relation, which in turn depends on the composition of planetary cores. Current studies suggest that super-Earth and mini-Neptune planets have iron-rich and hence largely Earth-like composition cores. This paper explores how non-zero planetary magnetic fields can decrease the expected mass-loss rates from these planets. These lower mass-loss rates, in turn, affect the location of the minimum of the radius distribution and the inferred chemical composition of the planetary cores. 
\end{abstract}




\section{Introduction}

Recent exoplanetary discovery missions have unveiled thousands of planets that reside close to their parent stars \citep[e.g.,][]{Morton2016}. The majority of these planets have radii in the range $R_p=1-4$~R$_\oplus$ \citep[e.g.,][]{Borucki2011,Fressin2013} and masses $M_p\lesssim 20$~M$_\oplus$ \citep{Howard2012,Weiss2014}. One overarching goal of exoplanet studies is to understand the distributions of planetary radii and masses in the observational sample. 

For those planets with measured masses and radii, we can use theoretical structure models to ascertain their properties. Many of the planets are consistent with being completely solid, with an approximately ``Earth-like'' composition (1/3 iron, 2/3 silicate rock) and do not possess a substantial atmosphere \citep{Dressing2015,Dorn2019}. However, many other planets have such low densities that they must be composed of a solid-core surrounded by a Hydrogen/Helium atmosphere with significant volume \citep[e.g.,][]{Jontof-Hutter2016}. Still other planets display intermediate densities, and could have a multitude of different compositions, ranging from a solid-core surrounded by a thin Hydrogen/Helium atmosphere to a planet containing large amounts of ice and water \citep[e.g.,][]{Rogers2010a}. 

One hypothesis that has gained significant observational support is that many of these close-in, low-mass planets originally formed with a composition that consisted of a solid core surrounded by a large H/He envelope. Due to the proximity of their host stars, however, these planetary atmospheres experienced severe photoevaporation driven by UV and X-ray radiation; this process was able to remove a significant fraction --- or in some cases all --- of their original H/He envelopes \citep{Owen2013,Lopez2013}. 

A key prediction from this scenario is that close-in, low-mass exoplanets evolving under the influence of photoevaporation follow two different pathways. Those planets with high irradiation levels, or with low mass cores, are completely stripped of their H/He atmosphere. In contrast, planets with lower irradiation levels, or with higher mass cores, retain a H/He envelope that contains roughly $\sim 1$\% of the core's mass \citep{Owen2013,Lopez2013}. As a $\sim 1$\% H/He envelope significantly increases the radius of a planet above the core surface, this latter evolutionary pathway leads to a gap in the distribution of exoplanet radii, where the location of the gap depends on the orbital period. This effect is often denoted as the ``evaporation valley''.  Using precise stellar radii determinations from the California-Kepler-Survey \citep{Petigura2017,Johnson2017}, \citet{Fulton2017} and \citet{Fulton2018} identified such a gap for planets with radii $R_P\sim1.8$~R$_\oplus$ and periods $P_{orb}<100$~days. \citet{vanEylen2018} used planets around stars with asteroseismically determined parameters to show that location of the gap (the characteristic planetary radius) decreases with increasing orbital periods as predicted by the photoevaporation model. 

Because the location of the gap in the radius distribution is identified observationally, and the core mass can be calculated from photoevaporation models, the orbital period at which a core of a given mass can be stripped depends on the mass-loss rate. As a result, using the observed location of the evaporation valley, coupled with a photoevaporation model, one can infer the composition of the planetary cores \citep{Owen2013,Owen2019}, and place constraints on where they formed within the protoplanetary disc. \citet{Owen2017}, \citet{Jin2018} and \citet{Wu2018}, using a photoevaporation model that assumes pure hydrodynamic outflow, derived a core composition that was predominately Earth-like, and contained little spread. This result indicates that core formation occurs inside the snow-line and perhaps favours in-situ formation models \citep[e.g.,][]{Hansen2012,Chatterjee2014,Lee2016,Jankovic2019}. However, this conclusion is dependent on the accuracy of the adopted mass-loss rates. Large-scale planetary magnetic fields, which remain unconstrained on exoplanets, can reduce the pure hydrodynamic mass-loss rates significantly. 
This present work explores the effects of magnetic fields on the mass-loss rates from close-in, low-mass planets. We then show how the presence of magnetic fields changes the estimates of the composition of planetary cores inferred from the location of the evaporation valley. 

\section{Theoretical Overview}

As detailed in Owen \& Wu (2013,2017), the origin of the evaporation valley is that the maximum timescale for evaporation (the ratio of the atmospheric mass to the mass-loss rate, $t_{\dot{m}}\equiv M_{\rm atm}/\dot{M}$) of a planet's atmosphere occurs when it roughly doubles the solid cores' radius. The timescale decreases for atmospheres smaller than this benchmark because the planetary radius is roughly constant (so that the mass-loss rate is roughly constant). The timescale also falls off for larger atmospheres as the planetary radius swells up rapidly with the addition of atmospheric mass (so that the mass-loss rate increases rapidly). For hydrogen/helium dominated atmospheres, this maximum timescale typically occurs when the atmosphere is a few percent by mass of the core. As a result, complete stripping of a core (independent of its initial atmospheric inventory) can occur when this maximum mass-loss timescale is shorter than the time available for mass-loss to occur. This latter time is typically $\sim 100$ Myr, the saturation timescale ($t_{\rm sat}$) for a star's high-energy output. { Thus, provided all cores accrete an envelope mass-fraction larger than that required to maximise the mass-loss timescale (for typical conditions this is true for cores more massive than $\gtrsim 1~$M$_\oplus$, see Section~3.2)},  the location of the evaporation valley is thus determined by the relation 
\begin{equation}
    t^{\rm max}_{\dot{m}}(M_p,a,t=t_{\rm sat}) \sim t_{\rm sat} \label{eqn:tmdot}\,,
\end{equation}
where $t^{\rm max}_{\dot{m}}(M_p,a,t=t_{\rm sat})$ is the maximum mass-loss timescale for a planet of mass $M_p$ with orbital separation $a$. 

\subsection{The Evaporation Valley as a Probe of Core Composition}

We can demonstrate how the observed position of the evaporation valley can be used to infer the core composition by approximately solving Equation~(\ref{eqn:tmdot}) to find the most massive planet that can be stripped at a given separation
(following \citealt{Owen2017}). We will solve the problem numerically in Section~3. In the following, we can parameterise the mass-loss rate using the energy-limited mass-loss law, where $\dot{M} = \eta \pi R_p^3L_{\rm HE}/4\pi a^2 G M_p$, with $L_{HE}$ the star's high-energy luminosity during the saturation phase, $R_p$ the planetary radius, and $\eta$ the mass-loss ``efficiency''. Given that the maximum mass-loss timescale occurs roughly for an atmospheric  mass fraction ($X_2$) that doubles the core radius (i.e., $R_p=2R_c$, \citealt{Owen2017}),  Equation~(\ref{eqn:tmdot}) becomes 
\begin{equation}
    \frac{GM_p^2X_2}{2 R_c^3} \approx \eta t_{\rm sat} \frac{L_{\rm HE}}{a^2} \,. \label{eqn:tmdot2}
\end{equation}
Since the planet's mass is completely dominated by the core, $M_p\approx M_c$, we can use a mass-radius relationship to convert the planet mass into a core radius through a mass-radius relationship. For demonstration purposes in this sub-section, we use $M_c \propto R_c^4$, so that the core's density $\rho_c$ is given by $\rho_c = \rho_{M_\oplus}(M_c/1\,{\rm M}_\oplus)^{1/4}$, with $\rho_{M_\oplus}$ the density of an Earth-mass core which depends on the core-composition. Dropping all constants, Equation~(\ref{eqn:tmdot2}) becomes 
\begin{equation}
    R_c \propto a^{-2/5}X_2^{-1/5} \eta^{1/5}\rho_{M_\oplus}^{-8/15}\,, 
    \label{eqn:valley}
\end{equation}
which specifies the position of the bottom of the evaporation-valley (i.e., the radii of the remaining stripped cores as a function of separation). For a given photoevaporation model (which determines $\eta(R_c,a)$) and an atmospheric structure model (which determines $X_2(R_c,a)$) the observed position of the valley in radius-separation space can be used to measure $\rho_{M_\oplus}$ and hence infer core composition.

Although the dependence and uncertainty of $X_2$ on various parameters is weak (\citealt{Owen2017}),
the dependence of the mass-loss efficiency $\eta$ is less well constrained, and is model dependent. As a result, the core-composition that one infers, based on the chosen form of $\eta$ and the observed location of the evaporation valley, is subject to a clear  model-dependent degeneracy. Smaller evaporation efficiencies $\eta$ can be balanced by lower-density cores to give the same valley position (see equation [\ref{eqn:valley}] and \citealt{Wu2018}).

\begin{figure*}
\centering
\includegraphics[width=\textwidth]{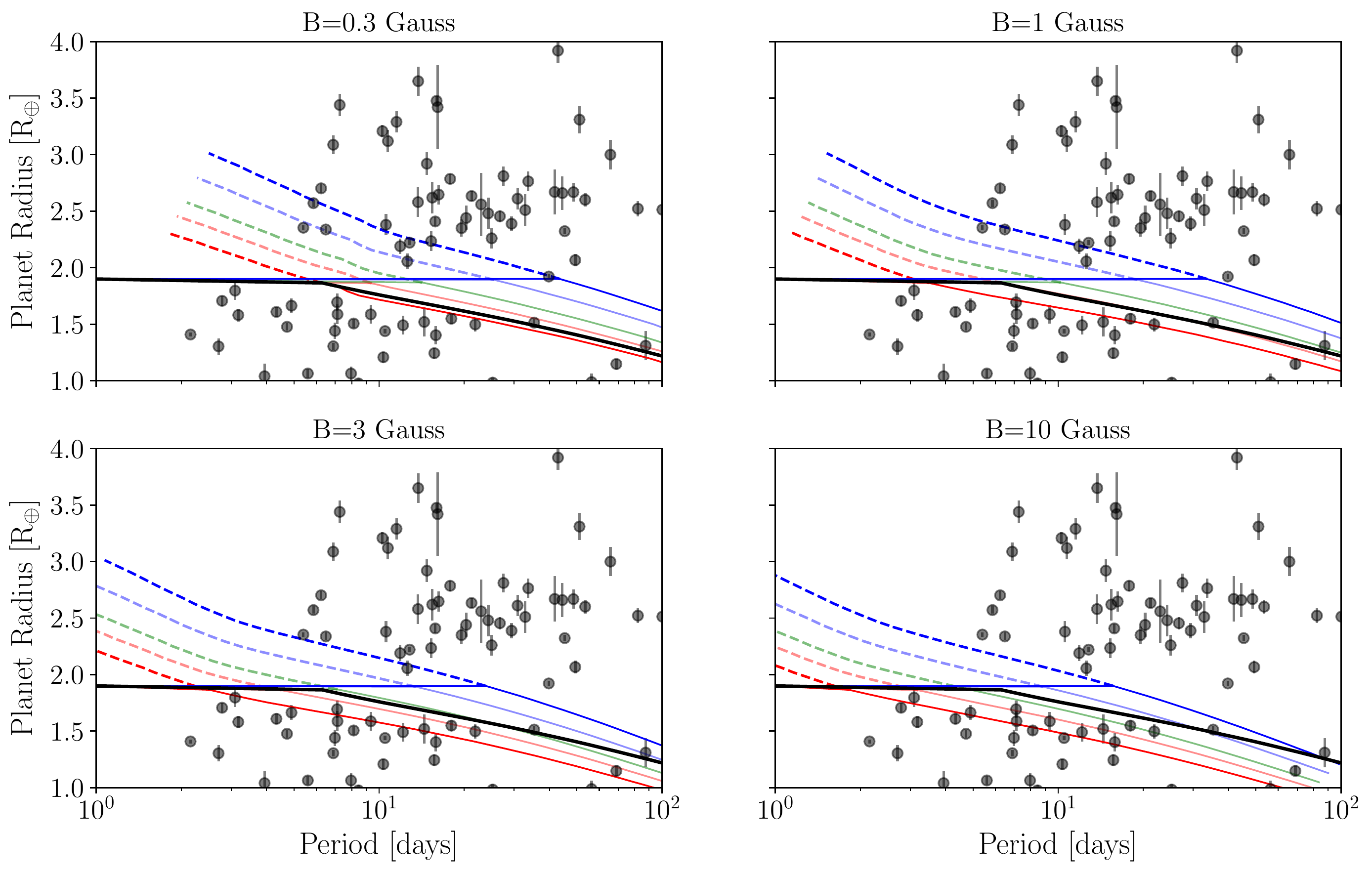}
\caption{The panels show the position of the lower boundary of the evaporation valley as a function of the core's composition for a magnetic field strength of 0.3 (upper left), 1.0 (upper-right), 3.0 (lower-left) and 10.0 (lower-right) gauss. The solid-black line shows the position for the pure hydrodynamic case ($B=0$). The coloured lines show core compositions of silicate rocks $+$ 1/3 iron (dark red), 15\% iron (light red), 15\% water-ice (light-blue) 1/3 water-ice (dark blue) and pure silicate rocks (green). Since the observed valley does not extend above $\sim 1.9$~R$_\oplus$ the calculated valley positions are shown as dotted lines above $1.9~$R$_\oplus$ and solid lines below. The data points are planets with precise parameters taken from Van Eylen et al. (2018).}\label{fig:panel_fig}
\end{figure*}
\begin{figure*}
\centering
\includegraphics[width=\textwidth]{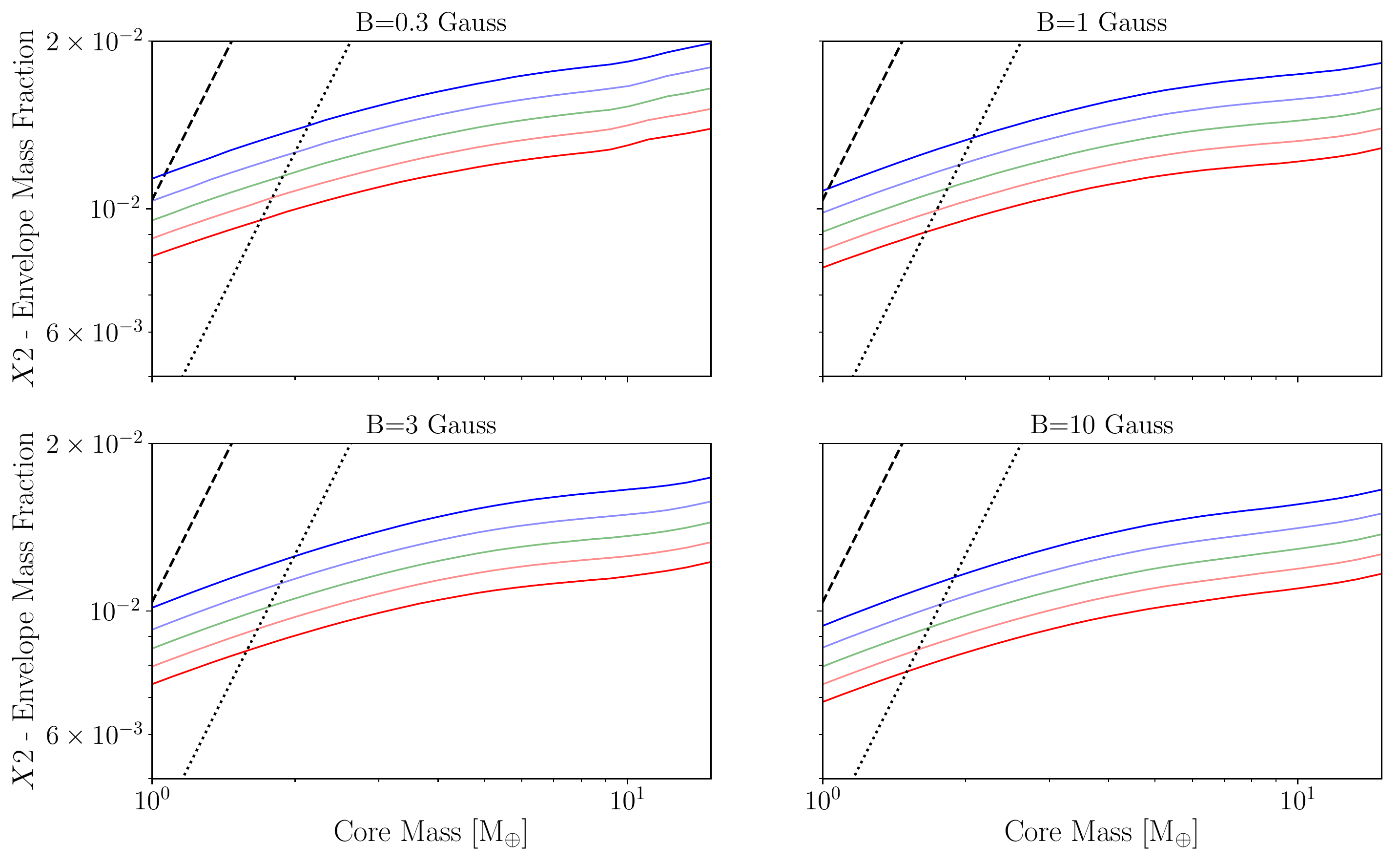}
\caption{The envelope mass fraction required to maximise the mass-loss timescale ($X_2$) as a function of core-mass for the models shown in Figure~\ref{fig:panel_fig}. The solid coloured lines show core compositions of silicate rocks $+$ 1/3 iron (dark red), 15\% iron (light red), 15\% water-ice (light-blue) 1/3 water-ice (dark blue) and pure silicate rocks (green). The dashed line show the envelope mass fraction a core of that mass would accrete in 1~Myr in a typical protoplanetary disc and the dotted line shows the envelope mass fraction that would be accreted in 1~Myr in a protoplanetary disc that is depleted in gas by a factor of 200; models taken from \citet{Lee2015}.}\label{fig:panel_fig2}
\end{figure*}
\subsection{Suppression of Mass-Loss by Planetary Magnetic Fields}

Close-in exoplanets could possess large-scale dynamo generated magnetic fields, either from their cores, or from their convective hydrogen/helium atmospheres. Since the evaporative outflow is at least partially ionized, it will be coupled to any large-scale planetary magnetic field \citep{Adams2011}. In the case of a dipole field, whose magnetic moment is aligned perpendicular to the planet's orbital plane, magnetic field lines near the planet's equator are likely to be closed. Such closed field lines --- by construction --- do not permit an outflow from low latitudes. This effect results in a ``dead-zone'' near the planet's equatorial region which is in magnetostatic balance \citep{Adams2011,Trammell2011,Trammell2014,DaleyYates2019}. The magnetic pressure drops rapidly (as $r^{-6}$) with increasing distance $r$, so that there will be some radius beyond which the magnetic pressure falls below the pressure of the planet's X-ray/UV heated atmosphere. Field lines originating at sufficiently high latitudes extend to larger radii and will be forced open, thereby allowing mass-loss from some fraction of the planetary surface. 

\citet{Owen2014} analytically estimated this fraction by computing magnetostatic balance of an isothermal atmosphere in the presence of a dipole magnetic field line. { This model was originally compared to simulations of more massive hot Jupiters, showing good agreement; however, the underlying physics of the model is independent of the underlying planet mass and is equally applicable to low-mass planets. } The fraction of the planet's surface that supports open field lines (and hence mass-loss) is then estimated by assuming that field lines which cross the equator in a region where the magnetic pressure is smaller than that of the isothermal atmosphere will be opened out by the outflow. This estimate gives an approximate solution for the fraction of the planetary surface that supports outflow, i.e., 
\begin{equation}
    F_{AP}\approx\frac{1}{2}\kappa^{-1/6}\exp\left(-\frac{\Phi_g}{6}\right)\,,\label{eqn:area}
\end{equation}
where $\kappa = B^2/(8\pi P_0)$ is a measure of the ratio of the magnetic pressure to thermal pressure at the planet's surface ($P_0$), $B$ is the dipole surface magnetic field strength, and $\Phi_g = G M_p/c_s^2R_p$ measures the depth of the gravitational potential well. Since not all of the planetary surface supports outflow, the mass-loss rate must be reduced in the presence of strong magnetic fields (large $\kappa$). 
For fixed planetary magnetic field strength, the parameter $\kappa$ depends on orbital separation through the pressure at the base of the outflow ($\propto a$ for EUV heating, \citealt{Owen2014}). However, since the exponents on $\eta$ and $\kappa$ are small in  Equations~(\ref{eqn:valley}) and (\ref{eqn:area}), the effect of magnetic fields on changing the slope of the valley is immeasurably small (unless $B$ varies strongly with core-mass\footnote{This possibility is (most likely) already ruled out, given that the valley slope is consistent with the standard model without magnetic fields (Van Eylen et al. 2018); see Section~\ref{sec:compare} for further discussion.}). 

Nonetheless, the presence of strong magnetic fields will act to reduce the overall mass-loss rate. For example, simulations have shown a field strength of 1~gauss is sufficient to reduce the mass-loss rates by an order of magnitude in hot-jupiters \citep[e.g.,][]{Adams2011,Owen2014,Arakcheev2017}. For the population of close-in super-Earths and mini-Neptunes,
an order of magnitude reduction in the mass-loss rates would make ice-rich cores, rather than iron-rich cores, consistent with the observed valley location. This change would obviously have important implications for the inferred formation location of these planets. We can analytically estimate the degeneracy between the core-composition and surface magnetic field strength by substituting into Equation~(\ref{eqn:tmdot2}) the result from Equation~(\ref{eqn:area}). The valley is observed to occur at some $R_c$ for a given $a$.  Neglecting the variation of $X_2$ and $\Phi_g$, and taking $P_0$ to be roughly independent of the magnetic field strength, we find the scaling relation 
\begin{equation}
    \rho_{M\oplus}\propto B^{-1/8}\quad{\rm for}\,\,B>B_{\rm min}\,, \label{eqn:B_approx}
\end{equation}
where $B_{\rm min}$ is the field strength below which magnetic fields are too weak to affect the outflow (i.e. $F_{\rm AP}(B\le B_{\rm min})=1$). Thus, stronger magnetic fields imply lower core densities, and ice-rich cores are only a factor of $\lesssim 2$ lower density than iron-rich ones.  

\section{Effects of Magnetic Fields}


The previous sections indicates that there is a degeneracy between the presence of magnetic fields on low-mass exoplanets and the core-composition inferred from the measured position of the evaporation valley. Here we solve for the this degeneracy for small, hydrogen/helium rich planets. { We emphasis we have no {\it a priori} knowledge of what magnetic field strengths to expect on exoplanets, other than picking values typical for Solar-System objects. A better way to think about this problem, which we discuss in Section~4, is using actual measurements of the core-composition of planets. By using follow-up mass-measurements, of planets below the evaporation valley, then one can, in combination with the position of the evaporation valley, make inferences about the magnetic field strengths of these planets.}

\subsection{Method}

As in Section~2, we solve for the lower-boundary of the evaporation valley by solving Equation~(\ref{eqn:tmdot}) as a function of orbital period, for different core compositions and magnetic field strengths. However, in this section we do it numerically. This calculations consists of two parts: First, a calculation of the planet's structure in order to find the radius and envelope mass-fractions at which the mass-loss timescale is maximised. Second, a calculation of the suppression of mass-loss by a magnetic fields. We consider our planet to consist of a solid-core surrounded by a hydrogen/helium envelope. The solid-core's mass-radius relationship is taken from the relations in \citet{Fortney2007}. In order to calculate the planetary radius and envelope mass-fraction at which the mass-loss timescale is maximised, we use the semi-analytic model of \citet{Owen2017}. In this model, the planet's envelope is approximated as an adiabatic interior (where we adopt $\gamma=5/3$), with an isothermal outer layer at the planet's equilibrium temperature.\footnote{Note that in this approach we solve the planetary structure equations, rather than using the simplified scaling relations.} 

For the mass-loss calculations we use the mass-loss rates and efficiencies from \citet{Owen2012}. In order to find the fraction of the planetary surface that permits outflow for a given field strength, we adopt the method of \citet{Owen2014}. Namely, the flow-profiles from \citet{Owen2012} provide the hydrodynamic variables along a streamline that connects the star and planet. Taking the planet to have a dipole field strength $B$ and a dipole direction that is perpendicular to the orbital plane, one can compare the pressure of the flow solutions to the magnetic pressure arising from an unperturbed dipole. Therefore, we can approximate the fraction of the planet's surface that permits outflow by finding the distance from the planet, along the streamline connecting the star and planet, where the flow pressure exceeds the magnetic pressure. The planet's dipole field structure can then be traced back from this distance to find the polar-angle at which it leaves the planet, and hence determine the fraction of the planetary surface that permits outflow. We then modify the mass-loss rate as $\dot{m}=F_{AP}\,\dot{m}_{HD}$, where $\dot{m}_{HD}$ are the mass-loss rates from the pure-hydrodynamic calculations of \citet{Owen2012}.

In all cases, surface magnetic field strengths are defined at the planet's photosphere. Recall that the maximum mass-loss timescale typically occurs when the planetary radii is twice that of the core. As a result, the magnetic field strength at the core's surface is approximately 8 times weaker at the planet's surface at the point where the mass-loss timescale is maximised. 

\subsection{Results and Discussion}\label{sec:compare}

Figure~\ref{fig:panel_fig} shows the position of the bottom of the evaporation-valley for core-compositions ranging from iron-rich and ``Earth-like'' to those which are ice-rich, for magnetic field strengths of 0.3, 1.0, 3.0 and 10.0 gauss. As expected, the presence of magnetic fields does not notably change the slope of the valley (although it does steepen for ice-rich cores, this is driven by changes in the mass-radius relation and mass-loss efficiency for icy-rich cores). However, it significantly shifts its position in the radius-period plane. Furthermore, the degeneracy between the inferred core-composition and assumed planetary magnetic field strength becomes apparent for magnetic field strengths $B\gtrsim 1$~gauss, where larger assumed magnetic field strengths require lower density and hence ice-rich cores.  

\begin{figure}
    \centering
    \includegraphics[width=\columnwidth]{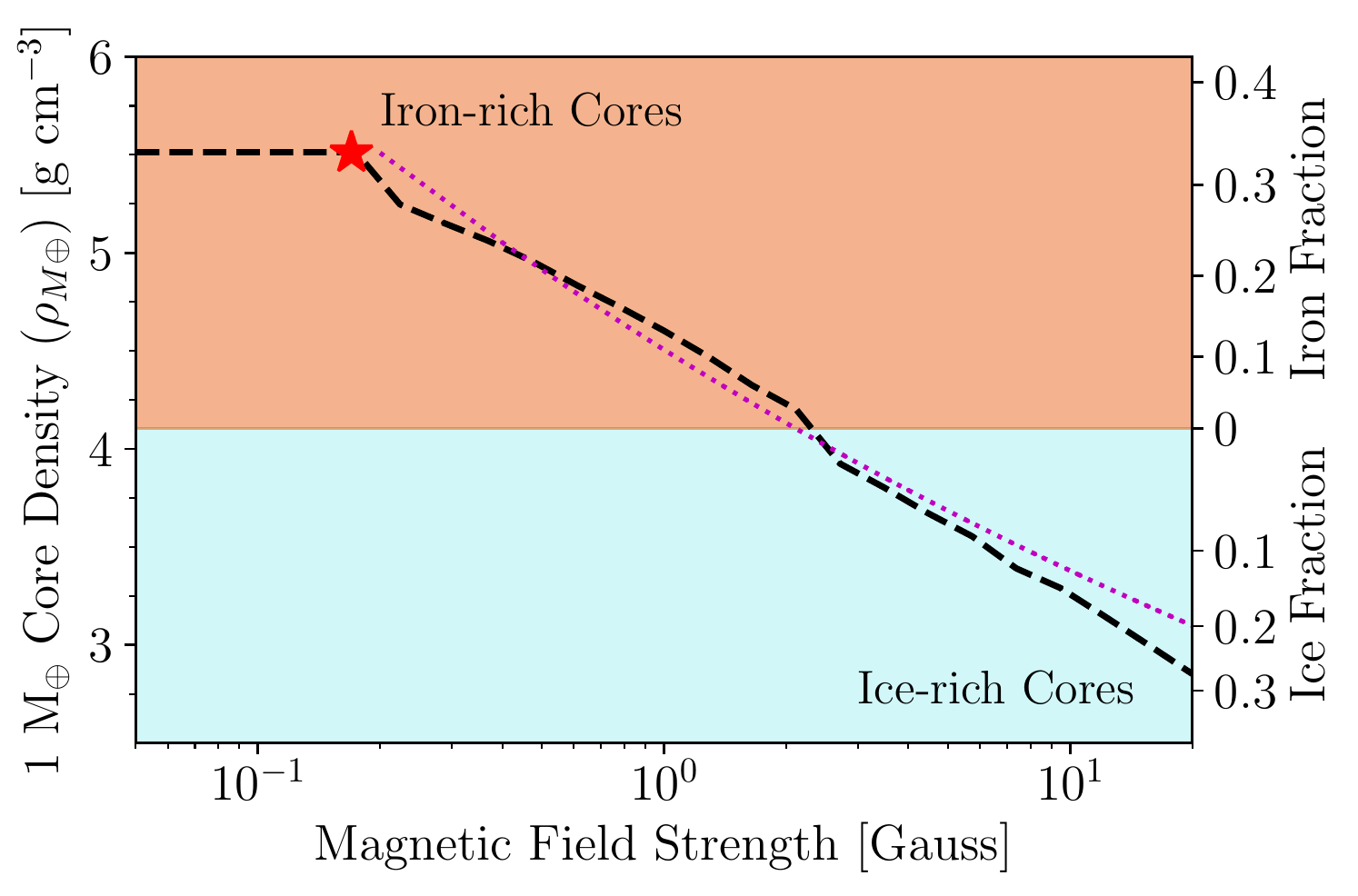}
    \caption{The density of an earth-mass core (and core-composition shown on the right-hand axis using the \citealt{Fortney2007} models) that would be inferred from the observed position of the evaporation valley as a function of the assumed planetary magnetic field strength. Higher assumed magnetic field strengths require lower density cores. The star shows the position of a 5~M$_\oplus$ Earth-like core with a 1~gauss field at the core-envelope boundary. The dotted line shows the approximate analytic scaling from Equation~\ref{eqn:B_approx}, which is in good agreement with the more detailed calculations. }
    \label{fig:comp_vs_B}
\end{figure}

{ As discussed in Section~2, our method implicitly assumes that all cores accrete sufficiently large hydrogen/helium envelopes so that the initial envelope mass fraction is larger than that required to maximise the mass-loss timescale ($X_2$). To check this assumption, we plot the value of the envelope mass fraction as a function of core-composition, core-mass, and magnetic field strength in Figure~\ref{fig:panel_fig2}. We see that for nominal conditions in protoplanetary discs, the core-accretion models of \citet{Lee2015} (shown as the dashed lines in Figure~\ref{fig:panel_fig2}) indicate that cores $\gtrsim 1$~M$_\oplus$ would typically accrete a sufficient hydrogen/helium envelope to satisfy our assumption (in agreement with other core-accretion models, which show typical envelope mass fractions $>1\%$, e.g., \citealt{Jin2014}). Even in the case of a highly gas depleted protoplanetary disc (dotted lines in Figure~\ref{fig:panel_fig2}) our assumption breaks down only below $\lesssim 2$~$M_\oplus$. Such cores would only be stripped at long periods, where we currently have limited data. As a result, our implicit assumption that cores accrete sufficient hydrogen/helium to initial have envelope mass fractions $>X_2$ is robust. This conclusion is also born out by the work of \citet{Wu2018}. In fitting the observed radius valley with  photoevaporation models, \citet{Wu2018} finds initial mass fractions $\gtrsim 1~$\%. If this assumption breaks down, however, the dependence of the evaporation-valley on the  accretion history of the cores cannot be ignored, meaning it may no longer provide a direct probe of core-composition.} 

The degeneracy between core-composition and magnetic field strength is more clearly seen in Figure~\ref{fig:comp_vs_B}, where we plot the inferred core density (shown for a 1~$M_\oplus$ core) as a function of the assumed surface magnetic field strength. The right side of the figure shows the composition expected for the core densities of the left side. This figure shows that ice-rich cores are preferred if planets have field strengths $B\gtrsim 3$~gauss, whereas iron-rich cores are indicated if planets have surface strengths $B\lesssim 3$~gauss. 

We also plot the surface magnetic field strength that a 5~M$_\oplus$ core with an Earth-like composition would have if the core had a 1~gauss field for an envelope that maximises the mass-loss timescale. This clearly indicates that for typical Solar-System terrestrial planet magnetic field strengths, if the magnetic fields were entirely generated in the core they would be too weak to affect the position of the evaporation valley and hence the core composition inferred. 

Therefore, a planetary field strength large enough to affect the inferred core composition from the evaporation valley would { likely} need to be dynamo-generated from within the hydrogen/helium atmosphere itself. Given there is no analogue of a planet that contains a few-percent of its mass in a hydrogen/helium atmosphere in our Solar-System, there are no direct examples of Solar-System bodies with measured field strengths for comparison. \citet{Christensen2009} provide a scaling relation, calibrated to the giant planets and stars, in order to turn convective heat flux into a magnetic field strength. Applying this relation to a several earth mass core with a few percent by mass hydrogen/helium envelope suggests that dipole field strengths of $\sim$5-10~gauss at an age of $\sim 100$~Myr are possible. However, we caution that these values come from extrapolating the \citet{Christensen2009} relation\footnote{In fact, we note that adopting the \citet{Christensen2009} relation in our full calculation predicts the incorrect slope for the evaporation valley.} far from where it was tested, and it's not clear that it can be blindly applied to a planet structure of our type. 

Another point of view on this problem can be found by looking at those planets below the evaporation valley with measured masses. Both \citet{Dressing2015} and \citet{Dorn2019} studied those planets with radii $R_p\lesssim 1.8$~R$_\oplus$ and well constrained masses, finding they are roughly consistent with a uniform, Earth-like and ice-free composition. This result suggests that the mass-loss from close-in mini-neptunes is not magnetically controlled, and that mass-loss rates are not suppressed by planetary magnetic fields. Taking this fact and turning it around would suggest that the hydrogen/helium envelopes of close-in mini-neputunes do not generate surface magnetic field strengths in excess of $\sim 0.3$~gauss during their youth and that the solid cores do not generate magnetic fields in excess of a few gauss at the core/envelope boundary. { Thus, future mass measurements of more planets below the evaporation-valley could be used to break the degeneracy between core-composition and magnetic field strength. This would allow insights into both planet formation and the strength of exoplanetary magnetic fields.

Clearly, since our comparisons are based on a statistical sample of exoplanets such conclusions only probe the bulk of the population. It is of course likely their is a distribution of exoplanetary magnetic field strengths.}

\section{Summary}

The observed location of the evaporation valley has been interpreted in favour of super-Earths and mini-neptunes possessing iron-rich, Earth-like composition cores. However, this inference is based on the mass-loss rates being accurate, and to date only pure hydrodynamic models have been used. It has been well established that strong planetary magnetic fields can suppress photoevaporative mass-loss, in some cases reducing the mass-loss rates by orders of magnitude. In this work, we have demonstrated that there is a degeneracy between the assumed surface planetary magnetic field strength and the core composition inferred from the observed evaporation valley. If the surface magnetic field strengths are in excess of $B\gtrsim 3$~gauss during the planet's first few 100 Myrs, then ice-rich rather than ice-free core would be inferred from the evaporation-valley. We have shown that magnetic fields generated within the solid core would be too weak by the time they reached the surface of the planet's hydrogen/helium atmosphere to suppress the mass-loss if they have magnitudes similar to terrestrial Solar-System bodies (e.g., $\sim$~gauss strengths at the core-envelope boundary). However, fields generated by a dynamo in the planet's convective hydrogen/helium atmosphere could suppress mass-loss and favour ice-rich cores. The fact that planets with precise mass values that reside below the evaporation valley are all consistent with an Earth-like composition suggests that dynamo generated fields in the hydrogen/helium envelopes of mini-neptunes are weak during their youth with surface field strengths $B\lesssim 0.3$~gauss.

\section*{Acknowledgements}
We are grateful to the referee for a report which improved our manuscript. JEO is supported by a Royal Society University Research Fellowship. JEO and FCA are grateful to The California Institute of Technology for hospitality during the beginning of this work. 





\bibliographystyle{mnras}
\bibliography{ref}


\bsp	
\label{lastpage}
\end{document}